\documentclass{PoS}

\title{Energy loss of a heavy quark in a hot QCD plasma}

 \ShortTitle{Energy loss of a heavy quark in a hot QCD plasma}

 \author{\speaker{S.P. Adhya}$ˆ{a}$, M. Mandal$ˆ{a}$,
 S. Sarkar$ˆ{b}$,  P. K. Roy$ˆ{a}$
 and S. Chattopadhyay$ˆ{a}$\\
\llap{$ˆa$} High Energy Nuclear and Particle Physics Division, Saha Institute
of Nuclear Physics,\\
1/AF Bidhannagar, Kolkata-700 064, INDIA\\
\llap{$ˆb$} Tata Institute of Fundamental Research, Homi Bhabha Road, Mumbai-400005, INDIA\\
E-mail: \email{souvikpriyam.adhya@saha.ac.in}}

\abstract{In this work we have studied the collisional energy loss of a heavy quark propagating through a high temperature QCD plasma consisting of both heavy and light quarks to leading logarithmic order in the Quantum Chromodynamics (QCD) coupling constant. The formalism adopted in this work shows a significant enhancement for the charm quark energy loss when the bath particles are also considered to be heavy in addition to light quarks. We know the running coupling constant is dependent on the momentum of the particles and the temperature of the system. Therefore, we have presented a comparison of the energy loss of the charm quark due to scattering with another heavy quark with constant and running coupling constant for different temperatures. The results show a substantial increase of the energy loss when compared to the fixed coupling case.
}

\FullConference{ 7th International Conference on Physics and Astrophysics of Quark Gluon Plasma\\
                 1-5 February , 2015 \\
                 Kolkata, India}

\def  \f    {\frac}

\def  \bef  {\begin{figure}}
\def  \eef  {\end{figure}}
\def  \be   {\begin{equation}}
\def  \ee   {\end{equation}}
\def  \ba   {\begin{array}}
\def  \ea   {\end{array}}
\def  \bea  {\begin{eqnarray}}
\def  \eea  {\end{eqnarray}}
\def  \beq  {\begin{eqnarray}}
\def  \eeq  {\end{eqnarray}}
\def  \nn   {\nonumber}
\def  \bd   {\begin{displaymath}}
\def  \ed   {\end{displaymath}}
\def  \bse  {\begin{subequations}}
\def  \ese  {\end{subequations}}
\def  \bwt  {\begin{widetext}}
\def  \ewt  {\end{widetext}}

\def  \ba   {{\bf{a_1}}}

\begin{document}

\section{Introduction}
The Quark Gluon plasma (QGP) formed in
 high energy experiments such as  RHIC and LHC provide us a scope for rich
investigative physics. One of the excellent probes of QGP are the formation of
jets in such \textit{magnum} collisions. The energetic partons during the
initial stages of the collision will rapidly propagate through the fireball
for a time period of a few $fm/c$ thus forming the jets. In the process, these
energetic partons will loose energy. The suppression pattern of high transverse
momentum hadrons, coined as "jet quenching"
provide an adequate tool to explain the creation of fireball inside the colliders. To understand quenching, one of the basic quantities is the rate
of energy loss per unit distance of a heavy quark produced in such a hot medium. The heavy quark, which in our consideration is the charm quark , losses energy via scattering with the partonic bath medium.

In this context, we recall that the energy loss of heavy quarks
have been carried  out by  Bjorken, Braaten, Peshier et. al.\cite{bjorken, thomaeloss1, thomaeloss2, peigne08a, peigne08b} where they
presented a theoretical framework to compute the losses. These heavy quarks are
the important analyzing tool of the primordial state of matter as they are
produced early in the timescale from the initial fusion of partons.
However, all these calculations did not consider the presence of charm quarks inside the medium which is quite relevant for the present temperatures of interest in the colliders.
 Thus, to establish the recent experimental data on the flow of $J/\psi$ at LHC with satisfactory
theoretical framework, one must consider the process of heavy quark energy loss
where the bath particles are also heavy and are at a equilibrium temperature. 

In the calculation of the elastic scattering process $QQ\rightarrow QQ$ in the $t-channel$, the contribution from the hard momentum transfer $q>q^*$ (where $Q=(\omega,\vec{q})$ is momentum transfer of the process $P+K\rightarrow P'+K'$ and $q^*$ is the intermediate momentum scale separating the soft and hard domains) is computed using a
tree level propagator for the exchanged gluon. On the other hand, the contribution from the
soft region $q<q^*$ is carried out using an effective gluon propagator
providing the necessary screening that cuts off the IR divergence at the Debye mass scale
$gT$. The scattering in the $s-channel$ do not contribute at leading log order and can therefore be ignored \cite{arnold00, arnold03,adhya14}.
Thus in this paper, we have presented a complete evaluation of the energy
loss
of an energetic heavy quark to the leading logarithmic order in the QCD coupling constant.
In addition, we have also presented results for running coupling and compared them with the fixed coupling case.
\section{Collisional energy loss ($QQ\rightarrow QQ$)}
The collision integral for a $2\rightarrow 2$ process can be expressed in the form,
\bea
\mathcal{C}[f_p]&=&\frac{1}{2E_p}\int \frac{d^3k}{(2\pi)^3 2E_k}
\frac{d^3p^{'}}{(2\pi)^3 2E_p'}\frac{d^3k^{'}}{(2\pi)^3 2E_k'}
 \delta f_p[PSF]\nn\\
&\times&(2\pi)^4 \delta^4(P+K-P'-K')
\frac{1}{2}\sum_{spin}{\cal| M |}_t^2,
\label{collision_term}
\eea
where, [PSF] denotes the phase space factor\cite{adhya14}. We have taken the scattering process $P+K\rightarrow P'+K'$ with appropriate energy and momentum labels. In our formalism, we have considered that the heavy quark losses energy by interacting with partons of the medium which include thermalized heavy quarks in addition to light quarks and gluons. In our case,
the heavy 
quark (of momentum $\vec{p}$) interacts with the thermal  heavy quarks of momentum $\vec{k}$.
From the expression of the collision integral, the heavy quark energy loss can be written as, 
\bea
\label{lossdef}
\Big(-\f{dE}{dx}\Big)_{QQ\rightarrow QQ} =\frac{1}{v_pE_p} \int_{p'}\int_k  
\int_{k'} (2\pi)^4 \delta^4(P+K-P'-K') \f{1}{2} \sum_{\rm spins} |{\cal M}|_t^2 
\omega  [PSF]  ,
\label{orieqn}
\eea
where, $\omega =(E_p -E_{p'})$ is the exchanged energy in an elastic scattering and $\int_{p'}$ 
denotes $d^3p'/(2E_{p'}(2\pi)^3)$ where we have considered both the quark 
and the anti-quark scatterings ($Q\bar Q\rightarrow Q\bar Q$). 
For such scattering process, the PSF can be re-calculated as \cite{abhee05},
\bea 
[PSF]=&&(f_{E_k}-f_{E_{k'}})\left [ 1 + \bar f_{q_0} - f_{E_{p'}} \right ]\nn\\
&&\simeq -\frac{df_{E_k}}{dE_k}q_0 \left[\frac{T}{q_0}+\frac{1}{2} \right ],
\eea
where, $\bar f_{q_0}$ is the boson distribution function and is given by $\bar f_{q_0}=(\exp(q_0/T)-1)^{-1}$ .
Firstly, we will focus on the calculation of the soft contribution of the charm quark energy loss. Hence we take into consideration the appropriate matrix amplitude with the HTL(Hard Thermal Loop) propagators \cite{adhya14},
\bea
\left(-\f{dE}{dx}\right)\Big|^{soft}_{QQ\rightarrow QQ}&=&\f{ g^4 C_F}{4v_p^2\pi^3 }\int  dq\int k
dk\f{1}{E_k}\int_{-v_p q}^{v_p
q}\omega^2\left(-\right)n_F'(E_k)d\omega\nn\\
&\times&\Big\{\mid \Delta_L(Q)\mid^2\f{E_k^2}{v_k }+\mid
\Delta_T(Q)\mid^2k^2\f{1}{2v_k }\Big[1-\f{\omega^2}{(v_k
q)^2}\Big]\Big[v_p^2-\f{\omega^2}{(v_k q)^2}\Big]\Big\}.
\label{dedxsoft1} 
\eea

The total contribution of heavy quark energy loss scattering off thermalized charms in the 
medium is obtained by adding both the  soft and hard contributions,
 \bea
 -\f{dE}{dx}\Bigg|_{QQ\rightarrow QQ}=-\f{dE}{dx}\Bigg|^{soft}_{QQ\rightarrow QQ}+
 -\f{dE}{dx}\Bigg|^{hard}_{QQ\rightarrow QQ}
 \label{tot_heavy_heavy}
 \eea
 It will be worthwhile to mention here that we can easily derive the corresponding hard part contribution to the energy loss by appropriate convertion of the propagators which will be bare ones instead of HTL propagators as in the domain of soft scattering. These calculations cannot be performed analytically and must be done numerically for complete evaluation of the energy loss of the charm quark with a thermalized charm in the medium. Further, it is easy to calculate the contribution of the energy loss for the processes ($Qq$ and $Qg$ scatterings) by suitable modification of the scattering species. 
\section{Results}
 \begin{figure}
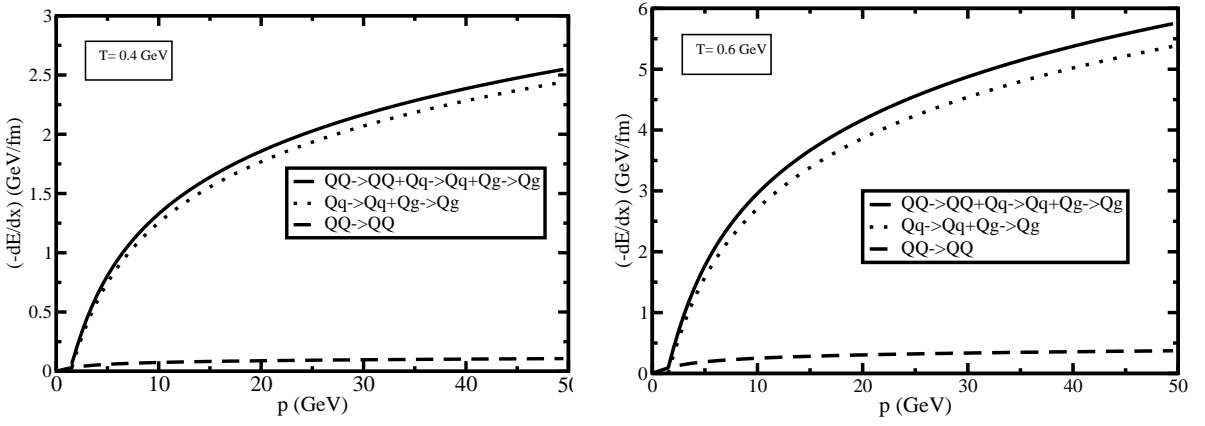

\centering
     \includegraphics[width=.5\textwidth]{latest_.4.eps}~~~~~
 \includegraphics[width=.5\textwidth]{latest_.6.eps}
  \caption{Energy loss $dE/dx$ of a charm quark as a function of its momentum for
 $T=0.4$ GeV (left panel) and $T=0.6$ GeV (right panel).}
 \label{fig2}
  \end{figure}
\begin{figure}
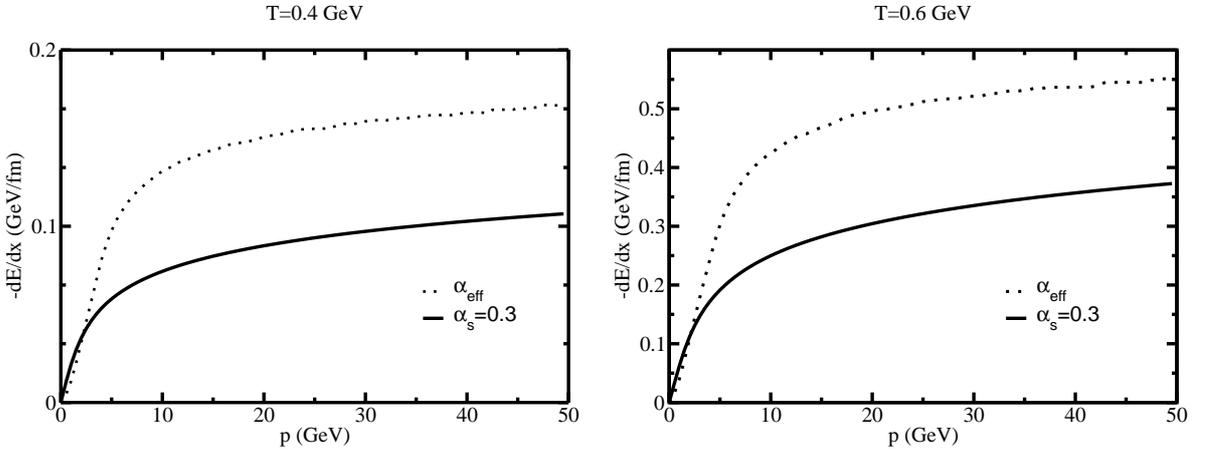

     \centering
     \includegraphics[width=.5\textwidth]{run_dok_.4.eps}~~~~~
     \includegraphics[width=.5\textwidth]{run_dok_.6.eps}
     \caption{Comparison of running coupling ($\alpha_{eff}$) on energy loss for 
 $QQ \rightarrow QQ$ scatterings at $0.4$ GeV (left panel) and at $0.6$ GeV (right panel) with the fixed value of the coupling constant ($\alpha_s$).}
     \label{fig4}
\end{figure}

We have made a comparison of the energy loss  in Fig.~(\ref{fig2}) between the two processes ($Qg\rightarrow Qg$ and $Qq\rightarrow Qq$) and $QQ\rightarrow QQ$ where the partonic bath consists of charm quark which is new.
We have presented plots of 
$(-dE/dx)$ with the momentum at temperatures of $0.4 GeV$ and $0.6 GeV$
respectively which are typical temperatures of QGP plasma produced at LHC.
In Fig.~(\ref{fig2}), we have presented comparison of  the energy losses of the charm quark (scattering with thermalized charm quarks) with the loss of the charm with the partonic bath particles.
In all our plots, we have found that the total energy loss of the heavy quark is substantially higher compared to the energy loss as calculated previously \cite{peigne08a,peigne08b} for the t-channel charm quark scattering  process. 

Secondly, we have plotted the loss of energy of charm quarks when the coupling is taken to be running and compared them with the fixed coupling case.
We have also found that the energy loss is quite higher when the coupling constant is taken be a function of momentum \cite{dokshitzer96}. 
\section{Summary}
In this work, we have considered the collisional energy loss due to elastic $QQ\rightarrow QQ$ scattering. In addition, similar scatterings of the heavy quark with the medium partons ($Qq\rightarrow Qq$ and $Qg\rightarrow Qg$) scatterings have also been taken into account to calculate the total energy loss of the charm quark. In our results, we have observed that with the increase in temperature, the charm quark loses energy significantly as compared to the case where the partonic medium is devoid of heavy quarks. 
It is also found that the energy loss is significantly higher when the coupling is taken to be running rather than a fixed one.
These observations can be attributed to the increase of number densities of the bath particles with temperature which in turn increases the interaction rate leading to enhanced charm quark energy loss as well which may provide theoretical insights into the study of  $v_2$ and $R_{AA}$ spectra of $J/\psi$.
\section*{Acknowledgment}
One of the authors (S. P. A.) would
like to thank UGC, India (Serial No. 2120951147) for
providing the fellowship during the tenure of this work.

\end{document}